\documentclass[preprint,showpacs,preprintnumbers,amsmath,amssymb]{revtex4}

\usepackage{graphicx}
\usepackage{dcolumn}
\usepackage{bm}

\begin{document}


\title{Observation of persistent photoconductivity in bulk Gallium Arsenide and Gallium Phosphide samples at cryogenic temperatures using the Whispering Gallery mode method}

\author{John G. Hartnett$^1$, David Mouneyrac$^{1,2}$, Jean-Michel Le Floch$^1$, Jerzy Krupka$^3$, Michael E. Tobar$^1$, D. Cros$^2$}
 \email{john@physics.uwa.edu.au}
\affiliation{$^1$School of Physics, University of Western Australia 35 Stirling Hwy Crawley 6009 W.A. Australia \\ $^2$XLIM UMR 6172 ­ Université de Limoges/CNRS ­ 123 Avenue Albert Thomas 87060 Limoges, Cedex, France \\ $^3$Institute of Microelectronics and Optoelectronics Department of Electronics Warsaw University of Technology Warsaw, Poland}

\date{\today}

\begin{abstract}
Whispering Gallery modes in bulk cylindrical Gallium Arsenide and Gallium Phosphide samples have been examined both in darkness and under white light at cryogenics temperatures $ \leq 50$ K. In both cases persistent photoconductivity was observed after initially exposing semiconductors to white light from a halogen lamp. Photoconductance decay time constants for GaP and GaAs were determined to be $0.900 \pm 0.081$ ns and $1.098 \pm 0.063$ ns, respectively, using this method.
\end{abstract}

\maketitle

\section{\label{sec:Intro}Introduction}
Persistent photoconductivity is extensively described in the literature\cite{Queisser1979, Queisser1984, Queisser1986, Schubert1985, Zardas1998, Lang1979}  as resulting from hole creation in a semiconductor where the energy gap is smaller than the energy of the illuminating light, thus freeing many new conduction electrons. Once the light source is extinguished recombination occurs. 

There are two main theoretical models.  One has the persistent photoconductivity resulting from the spatial separation of the photogenerated electrons and holes by macroscopic potential barriers due to band bending at planar surfaces, interfaces, junctions or around doping inhomogeneities. \cite{Queisser1979, Queisser1984, Queisser1986} This model has been extended with local inhomogeneities from defect clusters.\cite{Zardas1998} The second model has atomic scale  microscopic barriers at centers with large lattice relaxation. AlGaAs/GaAs structures are described \cite{Lang1979} with centers where the empty defect level lies above the minimum of the conduction band while the occupied level lies within the band gap. 

The main difference between microwave and DC techniques for the measurement of photoconductivity is associated with the electric field time-space distribution. In DC methods excess carriers are almost immediately removed from the material under test due to the presence of the DC electric field. \cite{Petritz1956} At microwave frequencies, however, the electric field and the carrier concentration alternate so one can observe recombination mechanisms in the volume of semiconductor, hence microwave techniques are superior. In order to obtain the highest measurement sensitivity though it is desirable to use measurement systems with the smallest energy dissipation. From this point of view resonance techniques are much better than the microwave transmission-reflection techniques--e.g. the microwave photoconductivity decay ($\mu$--PCD) method. \cite{Werner1986, Borrego1987}

In this paper we present the measurement of persistent photoconductivity in the semiconductors GaAs and GaP at cryogenic temperatures using the Whispering Gallery mode (WGM) method. We observe the time evolution of the Q-factor of chosen WG modes while the crystals are illuminated with white light, and after the light is extinguished.

\section{\label{sec:Measure}Measurement Technique}
The WGM method has become the most accurate method for measurements of the complex permittivity of extremely low-loss dielectric materials, both ceramic and crystalline. The method has been employed for very precise measurements of the permittivity and the dielectric losses of both isotropic and uniaxial anisotropic materials.  \cite{Baker-Jarvis1998, Krupka1997, Krupka1999a}  Very low-loss single-crystal materials including sapphire, ruby, Titanium doped sapphire, YAG, Chromium doped YAG, Calcium, Magnesium and Barium Fluoride, quartz, and others have been measured this way. \cite{Krupka1999b, Tobar1998, Hartnett2001, Tobar2001, Hartnett2002, Hartnett2004, Giordano2004, Jacob2006, Hartnett2006, Hartnett2007} The WGM technique has also been used to characterize the complex permittivity of semiconductors, including bulk monocrystalline Silicon, \cite{Krupka2006} Gallium Arsenide (GaAs) and Gallium Phosphide (GaP), \cite {Krupka2008, Hartnett2008} at microwave frequencies from cryogenic temperatures to room temperature.

The most effective way to eliminate cavity conductor losses in accurate dielectric permittivity and loss tangent measurements is to use high-order Whispering Gallery modes in cylindrical specimens of the material under test. In order to evaluate the complex permittivity of an isotropic sample a cylinder of the sample is loaded into a cylindrical copper cavity supported by central copper posts, excited by coupling in microwave energy via coaxial lines, and two WG modes are chosen that exhibit quasi-TE and quasi-TM field structures. Then the permittivity is calculated by matching the computed resonance frequencies with those measured. Finally from the permittivity resonance frequencies for several other modes are computed and compared with experiment, to check the validity of the mode identification. For a more thorough explanation of the method refer to references \cite{Baker-Jarvis1998, Krupka1997, Krupka1999a}.

Once the permittivity is found, the effective dielectric loss tangent is evaluated from the WG mode measured Q-factor using,
\begin{equation} \label{eqn:recipQ}
Q^{-1} = p_{\varepsilon}tan\delta + \frac{R_S}{G},
\end{equation}
where $tan\delta$ is the effective dielectric loss tangent of the crystal; $p_{\varepsilon}$ is the electric energy filling factor for the chosen WG mode in the sample under test (the ratio of the electric energy stored in the sample to the electric energy stored in whole resonator) defined as.   
\begin{equation} \label{eqn:ffactor}
p_{\varepsilon} = \frac{\iiint\limits_{Vd}\varepsilon_r|\textbf{E}|^2 dV}{{\iint\limits_{Vt}}\varepsilon_r(V)|\textbf{E}|^2 dV},
\end{equation}
where $Vt$ indicates the volume of the entire resonant structure and $Vd$ the volume of the sample under test.
 
The last term in (\ref{eqn:recipQ}) represents the conductor losses in the metal cavity walls. The parameters $R_S$ and $G$ are the surface resistance of the cavity walls and the geometric factor.  The latter may be evaluated from
\begin{equation} \label{eqn:Gfactor}
G = \omega\frac{\iiint\limits_{Vt}\mu_0|\textbf{H}|^2 dV}{{\iint\limits_S}|\textbf{H}_{\tau}| dS},
\end{equation}
where $\mu_0$ is the vacuum permeability, $S$ is the internal surface area of the cavity, $\omega$ is the angular frequency, and $\textbf{H}_{\tau}$ is the component of the magnetic field tangential to the internal surface of the cavity.

If a WG mode is chosen with  a sufficiently high enough frequency  the last term on the right-hand side of (\ref{eqn:recipQ}) may be neglected. Equation (\ref{eqn:recipQ}) assumes that there are no paramagnetic  impurities present in the sample.

In general, the complex permittivity of a semiconductor is given by
\begin{equation} \label{eqn:permittivity}
\varepsilon=\varepsilon_0\left(\varepsilon_r-j\varepsilon_r''-j\frac{\sigma}{\omega \varepsilon_0}\right) = \varepsilon_0\varepsilon_r(1-j \; tan\delta),
\end{equation}
where $\varepsilon_0$ is the vacuum permittivity, $\varepsilon_r$ and $\varepsilon_r''$ are the relative real and imaginary components of the 
permittivity of the semiconductor, and $\sigma$
is its conductivity. In this case the effective dielectric loss tangent of the semiconductor
is given by
\begin{equation} \label{eqn:tand}
tan\delta = tan\delta_d + \frac{\sigma_0 +\Delta \sigma}{\omega \varepsilon_0 \varepsilon_r}=\frac{\sigma_{0eff}}{\omega \varepsilon_0 \varepsilon_r}+\frac{\Delta \sigma}{\omega \varepsilon_0 \varepsilon_r},
\end{equation}
where $tan \delta_d$ is the dielectric loss tangent associated
with pure dielectric loss mechanisms (e.g., electronic and ionic polarization), $\sigma_0$ is the conductivity in the semi-conductor in darkness, and $\Delta\sigma$ is the excess conductivity associated with free carriers generated by illumination from a light source.

For doped and intrinsic semiconductors having energy gaps of order 1 eV or less, the dominant loss mechanism, for excitation frequencies as high as microwaves and at temperatures that exceed the activation energy of the dopands, is related to the conductivity associated with free charge carriers.  For such materials, their dielectric loss tangent can be represented by the expression on the right-hand side of (\ref{eqn:tand}). The effective conductivity $\sigma_{0eff}$ includes dielectric losses. 

Measurements were performed on cylindrical samples of pure GaAs and GaP. \cite{Krupka2008} The GaAs sample has a diameter of $25.39 \pm  0.01$ mm and a height of $6.25 \pm  0.01$ mm. The GaP sample has a diameter of $48.12 \pm  0.03$ mm and a height of $5.00 \pm  0.01$ mm. These samples have been mechanically polished on all surfaces. The internal dimensions of the cylindrical copper cavities used in WGM measurements are as follows. The cavity used with GaAs sample has a diameter of 34.6 mm and a height of 25.8 mm. The cavity used with the GaP sample has a diameter of 60.0 mm and a height of 43.0 mm. 

The resonators were coupled to a vector network analyzer (VNA) via coaxial cables. They were located in an evacuated chamber and cooled on the cold-finger of a single-stage cryocooler. WG mode families were identified at cryogenic temperatures. The mode frequency, loaded Q-factor and couplings were calculated from the measured VNA S-parameters and then the unloaded Q-factor was determined. Table I shows data for two mode families identified in the different samples. For the following analyses only the highest Q-factor modes in each sample were selected, where the Q-factors are not limited by paramagnetic losses or the cavity walls. They are at the limit of the dielectric loss tangent plus some contribution from $\sigma_0$ the conductivity in the semi-conductor in darkness.

The modes have been labeled N(S)X-$m$ according to Krupka \cite{Krupka1999b} where N or S respectively indicate whether the magnetic field is anti-symmetric or symmetric with respect to the plane of the coupling loops. (The coupling loops are located at the midpoint on the cylindrical wall of the cavity, on opposite sides.) The parameter X indicates the order of increasing value of the frequency within a mode family and $m$ is the azimuthal mode number. The mode electric energy filling factors were calculated using (\ref{eqn:ffactor}). The values of electric filling factors were calculated once the relative permittivity was known. 

\section{Photoconductivity}
Each sample was illuminated with light that was conducted into the cryogenic resonator via optical fibers. The energy gap for GaAs and GaP are 1.424 eV and 2.26 eV respectively. \cite{GaAs, GaP} Therefore we expect threshold wavelengths of the illuminating light, to excite photoconduction, to be 871 nm (infrared (IR)) and 549 nm (green) respectively. A white light source emitting about 10 mW of total light was used in both cases. This source was quite hot and emits significant IR. 

Under the white light the microwave losses in GaP reduced the mode Q-factor by about 33\%.  In the GaAs the microwave losses were so significant that the mode almost disappeared completely (from the VNA screen) when the light was on. Figure 1 shows the time dependence of the microwave losses (or $Q^{-1}$ specified by (\ref{eqn:recipQ})) of two N1 modes in GaAs at 36 K, both initially in darkness and after illumination. The light remained on until the mode could no longer be seen. After it was switched off the unloaded Q-factor for the mode in question was recorded as a function of time. 

In both modes the microwave losses do not return to the values recorded before any illumination at all. Clearly the losses remain indefinitely much higher. It follows from (\ref{eqn:recipQ})  and (\ref{eqn:tand})  that 
\begin{equation} \label{eqn:tand2}
\Delta \sigma = \left(\frac{1}{p_{\epsilon} Q}-\frac{1}{p_{\epsilon} Q_0}\right)\omega \varepsilon_0 \varepsilon_r,
\end{equation}
where the microwave losses $(p_{\epsilon}Q_0)^{-1}$ may be neglected when the photoconductivity ($\Delta \sigma$) generated by the illumination process is such that $Q^{-1} \gg Q_0^{-1}$.  Here  $Q_0$ is the Q-factor of the mode before any illumination at all.

A continuous data acquisition system was implemented that would record both the loaded Q-factor and frequency of the resonant mode on  the VNA.  Couplings were set sufficiently low that the loaded Q-factor is approximately equal to the loaded Q-factor. With both GaAs and GaP samples the white light source was switched on, data recorded and then switched off and data again recorded. This is shown in Figs 2 and 4.  

From the data of Fig. 1, $p_{\varepsilon}$ and Q-factor from Table I for the 18.949 GHz N1-13 mode in GaAs, the semiconductor conductivity was calculated using (\ref{eqn:tand2}). The results are shown in Fig. 2 at 50 K and in Fig. 3 at both 50 K and 36 K, as a function of time (in seconds) on a logarithmic  axis. From Fig. 2 it is seen that the photoconductance decays quickly at first then experiences a lower rate of decay. It is possible that the initial two points in Fig. 2 are due to thermal effects on the sample as the light is switched off, but the frequency response\cite{Hartnett2008} in both GaP and GaAs were independent of how quickly we disconnected the fiber. At most thermal effects are small compared to the results shown here.

The data, as shown in Fig. 3, follow a natural logarithmic decay but eventually the excess conductivity becomes constant at times above $3 \times 10^{5}$ s at 50 K  and above $10^{6}$ s at 36 K. Note the excess conductivity is  greater at 36 K than at 50 K.

From the rectangular spatial Fermi distribution model of Queisser and Theodorou \cite{Queisser1986} (their Eq. (11)) the decaying photoconductivity can be expressed as
\begin{equation} \label{eqn:sigma}
\Delta \sigma(t) = \Delta \sigma_0-  A\, ln\left(1+(t/\tau_0)\right),
\end{equation}
where $\Delta \sigma_0$ is the excess conductivity at $t=0$ just when the light is switched off; $A \propto \frac{1}{2}a Z$ is a measure of the photoconductance lost within a time $\tau_0$; $a$ is the Bohr radius; $Z$ is the density of hole-capturing traps in the material. This model fits very well to the data of Fig. 2 up to $t = 4 \times 10^5$ s. For $t > 4 \times 10^5$ the excess conductivity remains constant at both 36 K and 50 K in GaAs; the modeled curve falls below.

The best fit values using the above model, using the 50 K data, results in $\tau_0 = (10.98 \pm 0.63) \times 10^{-10}$ s, consistent with luminescence experiments. \cite{Queisser1986}  Table II gives the resulting best fit values for $\Delta \sigma_0$ and $A$ at the two temperatures measured.

The GaP sample was also temperature stabilized in the cryostat at 50 K and the frequency and Q-factor of the 11.544 GHz S1-12 mode was continuously recorded before and after the white light source was switched on then off. 

Figure 4 shows the excess conductivity calculated using (\ref{eqn:tand2}). The times of switching are indicated. In this case the process was repeated a number of times and so prior to ``light ON'' in Fig. 4 the sample was in the persistent current condition. In the GaP sample the persistent photoconductivity was realized within about 3000 s of switching the light off. See Fig. 4 between $7 \times 10^3$ s and $1.1 \times 10^4$ s. In this plot the excess conductivity in darkness (by definition) before any light was switched on at all is zero.  

The small apparent rise in photoconductivity in Fig. 4  before ``light ON''  is due to a frequency shift which resulted from the vacuum pump being recently turned on again. Other runs don't show this. However this data run shows best the second order response system seen in both the mode bandwidth and frequency, here circled.\cite{Hartnett2008} 

In GaP we found that the model (\ref{eqn:sigma}) also fits very well to the data of Fig. 4, after the light was switched off. The result is shown as the (red) broken curve in Fig. 5 where the fit has been applied to the measured (gray) data points.  Table II lists the resulting best fit values for $\Delta \sigma_0$ and $A$ at 50 K. The best fit decay time was found to be $\tau_0 = (9.00 \pm 0.81) \times 10^{-10}$ s in this case.

\section{Discussion and Conclusion}
Whispering Gallery modes in bulk cylindrical Gallium Arsenide and Gallium Phosphide samples have been examined both in darkness and under white light at cryogenics temperatures. The semiconductors were temperature stabilized at temperatures $ \leq  50$ K. In both cases persistent photoconductivity was observed after initially exposing the samples to white light from a halogen lamp. Also we observed the change in permittivity under light and dark conditions which is related to the polarization state of the semiconductor. This has been reported previously. \cite{Hartnett2008} To gather the data for the excess conductivity in GaAs however has taken much longer than was necessary for the analysis related to changes in permittivity under illumination.

It was observed that GaAs takes a very long time to achieve constant conductivity, well above that observed in darkness before photo-electrons are excited from the bulk. The very long decay time of the excess conductivity agrees very well with the rectangular spatial Fermi distribution model of Queisser and Theodorou. \cite{Queisser1986}  GaP on the other hand very quickly achieves persistent photoconductivity after the illuminating light source is switched off. This also is well above the conductivity in darkness before photoconductivity is induced. 

\section{Acknowledgment}
This work was supported by the Australian Research Council. We would like to thank the Institute of Electronic Materials Technology, 133 Wolczynska St, 01-919 Warsaw, Poland for supplying the semiconductor samples.

\newpage

\begin{table}[ph]
\begin{center}
\caption{\label{tab:table1}Whispering Gallery mode data: GaAs at 41 K (top); GaP at 50 K (bottom) }
\begin{tabular}{lcccc}
\hline\hline
&$Freq. \, [GHz]\,\,\,$ &$Q$-factor 	&$\,\,\,\,\,\,\,\,\,p_{\varepsilon}\,\,\,\,\,\,\,\,\,$ &$\,\,\,\,Label\,\,\,\,$ \\
\hline
& 12.502347 & 7.18 $\times \, 10^5$		& 0.96323 	& N1-7  \\
& 13.546221 & 7.33 $\times \,10^5$ 	& 0.97097 	& N1-8 \\
& 14.606873 & 8.57 $\times \,10^5$		& 0.97528 	& N1-9  \\
& 15.680292 & 9.07 $\times \,10^5$ 	& 0.97989 	& N1-10 \\
& 16.763338 & 9.79 $\times \,10^5$ 	& 0.98374  	& N1-11 \\
& 17.853708 & 9.99 $\times \,10^5$ 	& 0.98527 	& N1-12\\
& 18.949572 & 1.10 $\times \,10^6$ 	& 0.98577 	& N1-13 \\
& 20.049624 & 1.00 $\times \,10^6$ 	& 0.98683  	& N1-14 \\
& 21.153240 & 9.00 $\times \,10^5$ 	& 0.98845  	& N1-15 \\
& 22.259175 & 7.95 $\times \,10^5$ 	& 0.99023  	& N1-16 \\
& 23.366946 & 6.00 $\times \,10^5$ 	& 0.99177  	& N1-17 \\
\hline
& 10.910025 & 1.68 $\times \,10^5$		& 0.96247 	& S1-11 \\
& 11.544811 & 1.80 $\times \,10^5$ 	& 0.96672  	& S1-12 \\
& 12.170065 & 						 	& 0.97105  	& S1-13  \\
& 12.804506 & 							& 0.97434 	& S1-14  \\
& 13.427628 & 1.77 $\times \,10^5$ 	& 0.97584 	& S1-15 \\
& 14.054495 &  						 	& 0.97635 	& S1-16 \\
& 14.677084 & 						 	& 0.97707		& S1-17\\
& 15.294796 & 1.80 $\times \,10^5$	 	& 0.97919 	& S1-18\\
\hline
\end{tabular}
\end{center}
\end{table}

\begin{table}[ph]
\begin{center}
\caption{\label{tab:table2}Conductivity data: GaAs (top two rows), GaP (bottom row)  }
\begin{tabular}{lccc}
\hline\hline
&\,\,T[K]\,\, &$\,\,\Delta \sigma_0 \, [Sm^{-1}]\,\,$ &$\,\,A\, [Sm^{-1}]\,\,$  \\
\hline
&50 K & $(1.71 \pm 0.04) \times 10^{-3}$			&$(4.11 \pm 0.11) \times 10^{-5}$ \\
&36 K & $(1.78 \pm 0.05) \times 10^{-3}$ 		&$(4.11 \pm 0.15) \times 10^{-5}$\\
\hline
&50 K & $(8.08 \pm 0.74) \times 10^{-5}$			&$(2.59 \pm 0.03) \times 10^{-6}$ \\
\hline
\end{tabular}
\end{center}
\end{table}

\newpage

Figure Captions:

Figure 1: Log-linear plot of $Q^{-1}$ of two N1 modes in GaAs measured at 36 K, as a function of time after the illumination is turned off. The filled squares (top) represent the 12.50 GHz N1-7 mode and the open circles (bottom) represent the 18.949 GHz N1-13 mode. Error bars are $\pm 5\%$ of the measured data.

Figure 2: Photoconductivity $\Delta \sigma$ at 50 K in GaAs at 18.949 GHz before the light was switched on, then switched on for about 200 s, then after it was switched off.  

Figure 3: Photoconductivity $\Delta \sigma$ at both 50 K (open circles -- bottom curve -- same data as  in Fig. 2) and 36 K (solid circles -- top curve), in GaAs at 18.949 GHz.  Error bars follow from the $\pm 5\%$ error on the $Q$ measurements. 

Figure 4: The black dots are the smoothed data for $\Delta \sigma$ at 11.544 GHz derived from the S1-12 mode in GaP measured at 50 K, as a function of time, both before and after the illumination with white light source. The gray connected dots are the measured data. The noise is due to the errors in the bandwidth determination.

Figure 5: Expanded section of data taken from Fig. 4. The black dots are the smoothed data for $\Delta \sigma$ at 11.544 GHz derived from the S1-12 mode in GaP measured at 50 K, after the light source was switched off. The gray connected dots are the measured data. The (red) broken curve is fitted to the measured (gray) data.


\begin{thebibliography}{99}
\bibitem{Lang1979} D.V. Lang, R.A. Logan, and M. Jaros, Phys. Rev. B, \textbf{19}, 1015 (1979).
\bibitem{Queisser1986} H.J. Queisser and D.E. Theodorou, Phys. Rev. B, \textbf{33}, 4027 (1986).
\bibitem{Queisser1979} H.J. Queisser and D.E. Theodorou, Phys. Rev. Lett., \textbf{43}, 401 (1979).
\bibitem{Queisser1984} H.J. Queisser and D.E. Theodorou, Solid State Commun., \textbf{51}, 875 (1984)
\bibitem{Schubert1985} E.F. Schubert, A. Fischer, and K. Ploog, Phys. Rev. B, \textbf{31}, 7937  (1985).\bibitem{Zardas1998} G.E. Zardas, D.E. Theodorou, P.C. Euthymiou, Ch. I. Symeonides, F. Riesz, and B. Szentpall, Solid State Commun., \textbf{105}, 77 (1998).
\bibitem{Petritz1956} R.L. Petritz, Phys. Rev. \textbf{104}, 1508 (1956).
\bibitem{Borrego1987} J. M. Borrego, R. J. Gutmann, N. Jensen, O. Paz, Solid-State Electronics, \textbf{30}, 195 (1987).
\bibitem{Werner1986} A. Werner, M. Kunst, G. Beck, J. Christen,  and G. Weimann, Phys. Rev. B \textbf{34}, 5918 (1986). 
\bibitem{Baker-Jarvis1998}	J. Baker-Jarvis, R. G. Geyer, J. H. Grosvenor Jr, M. D. Janezic, C. A. Jones, B. Riddle, C. M. Weil, and J. Krupka, IEEE Trans. Dielectric Electric. Insul., \textbf{5}, 571 (1998).
\bibitem{Krupka1997} J. Krupka, K. Derzakowski, A. Abramowicz, M. E. Tobar, and R. G. Geyer,  Proceedings of IEEE MTT Int. Micr. Sym. Digest, Denver, (1997).
\bibitem{Krupka1999a}	J. Krupka, K. Derzakowski, M. E. Tobar, J. G. Hartnett, and R. G. Geyer, Meas. Sci. Technol., \textbf{10}, 387 (1999).
\bibitem{Krupka1999b}	J. Krupka, K. Derzakowski, A. Abramowicz, M. E. Tobar, and R.G.  Geyer, IEEE Trans. on MTT, \textbf{47}, 752 (1999).
\bibitem{Hartnett2001}	J. G. Hartnett, M. E. Tobar, and J. Krupka,  J. Phys. D: Appl. Phys., \textbf{34}, 959 (2001).
\bibitem{Hartnett2002}	J. G. Hartnett, A. N. Luiten, J. Krupka, M. E. Tobar, and P. Bilski,  J. Phys. D: Appl. Phys., \textbf{35}, 1459 (2002).
\bibitem{Hartnett2004}	J. G. Hartnett, A. C. Fowler, M. E. Tobar, and J. Krupka,  IEEE Trans. on UFFC, \textbf{51}, 380 (2004).
\bibitem{Giordano2004} V. Giordano, J.G. Hartnett, J. Krupka, Y. Kersalé, P.Y. Bourgeois, M.E. Tobar, IEEE Trans. on UFFC, \textbf{51}, 484 (2004).
\bibitem{Hartnett2006} J.G. Hartnett, M.E. Tobar, E.N. Ivanov and J. Krupka, IEEE Trans. on UFFC, \textbf{53}, 34 (2006).
\bibitem{Hartnett2007} J.G. Hartnett, M.E. Tobar, J-M. le Floch, J. Krupka and P-Y. Bourgeois,  Phys. Rev. B, \textbf{75}, 024415 (2007).
\bibitem{Jacob2006}	M. V. Jacob, J. G. Hartnett, J. Mazierska, J. Krupka, and M. E. Tobar,  Cryogenics, \textbf{46}, 730 (2006).
\bibitem{Tobar1998}	M. E. Tobar, J. Krupka, E. N. Ivanov, and R. A. Woode,  J. of Appl. Phys, \textbf{83}, 1604 (1998).
\bibitem{Tobar2001}	M. E. Tobar, J.G. Hartnett, E.N. Ivanov, D. Cros, P. Blondy, IEEE Trans. on Instrum. \& Meas., \textbf{50}, 522, (2001).
\bibitem{Krupka2006} J. Krupka, J. Breeze, A. Centeno, N. Alford, T. Claussen, and L. Jensen, IEEE Trans. on MTT, \textbf{54}, 3995 (2006).
\bibitem{Krupka2008} J. Krupka, D. Mouneyrac, J.G. Hartnett and M.E. Tobar, IEEE Trans. on MTT, \textbf{56}, 1201 (2008).
\bibitem{Hartnett2008} J.G. Hartnett, D. Mouneyrac, J-M Le Floch, J. Krupka, M.E. Tobar, D. Cros, Appl. Phys. Lett. \textbf{93}, 062105, (2008). 
\bibitem{GaAs} www.ioffe.rssi.ru/SVA/NSM/Semicond/GaAs
\bibitem{GaP} www.ioffe.ru/SVA/NSM/Semicond/GaP
\end{thebibliography}
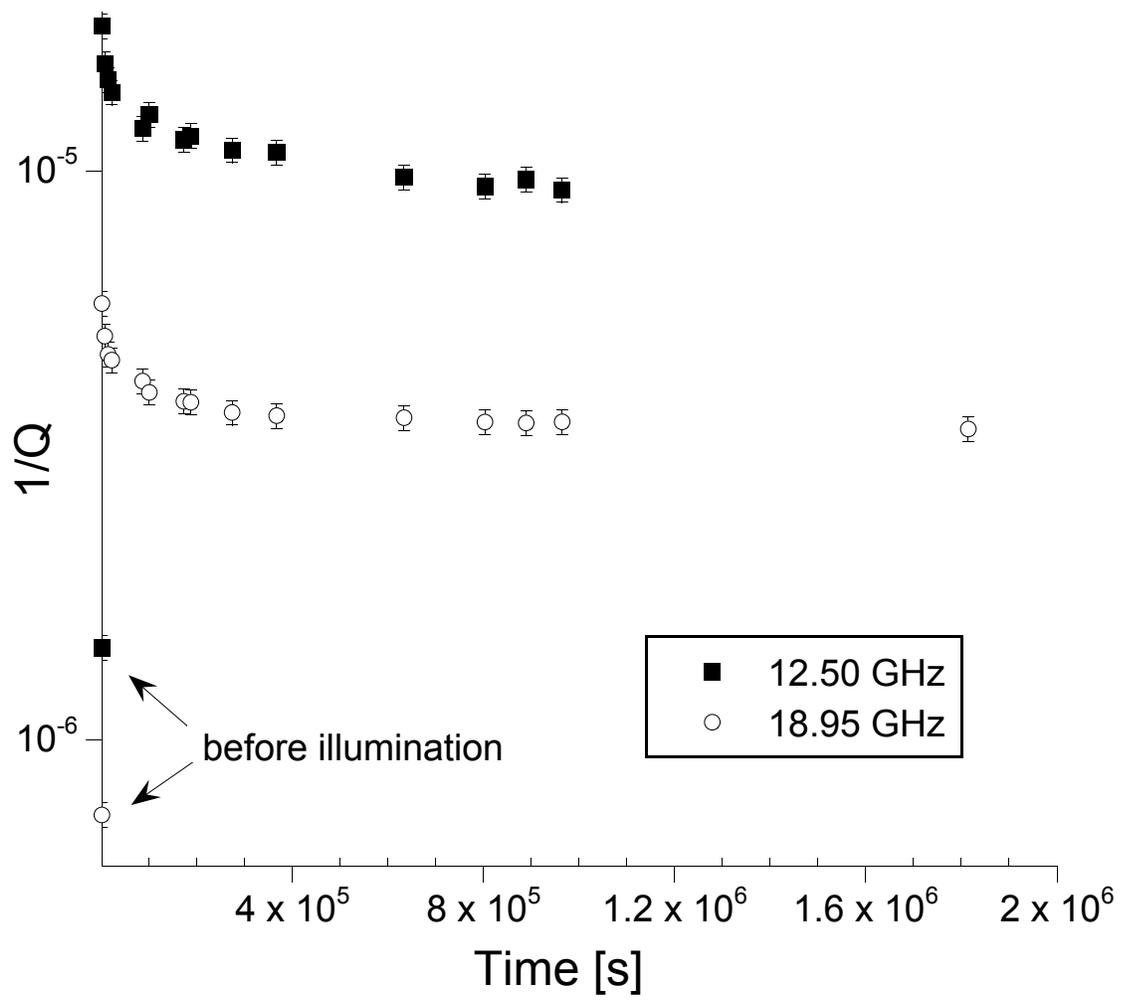
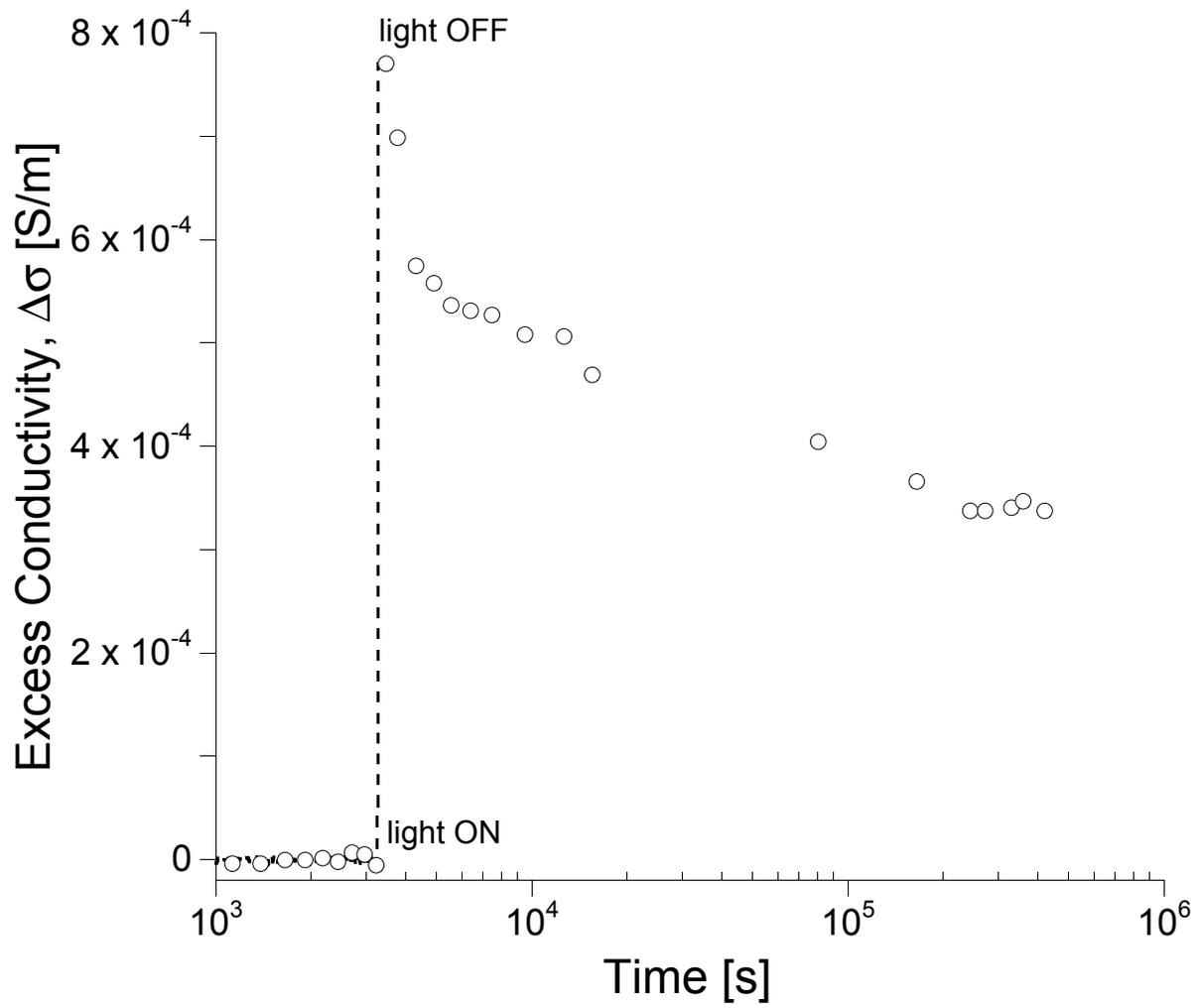
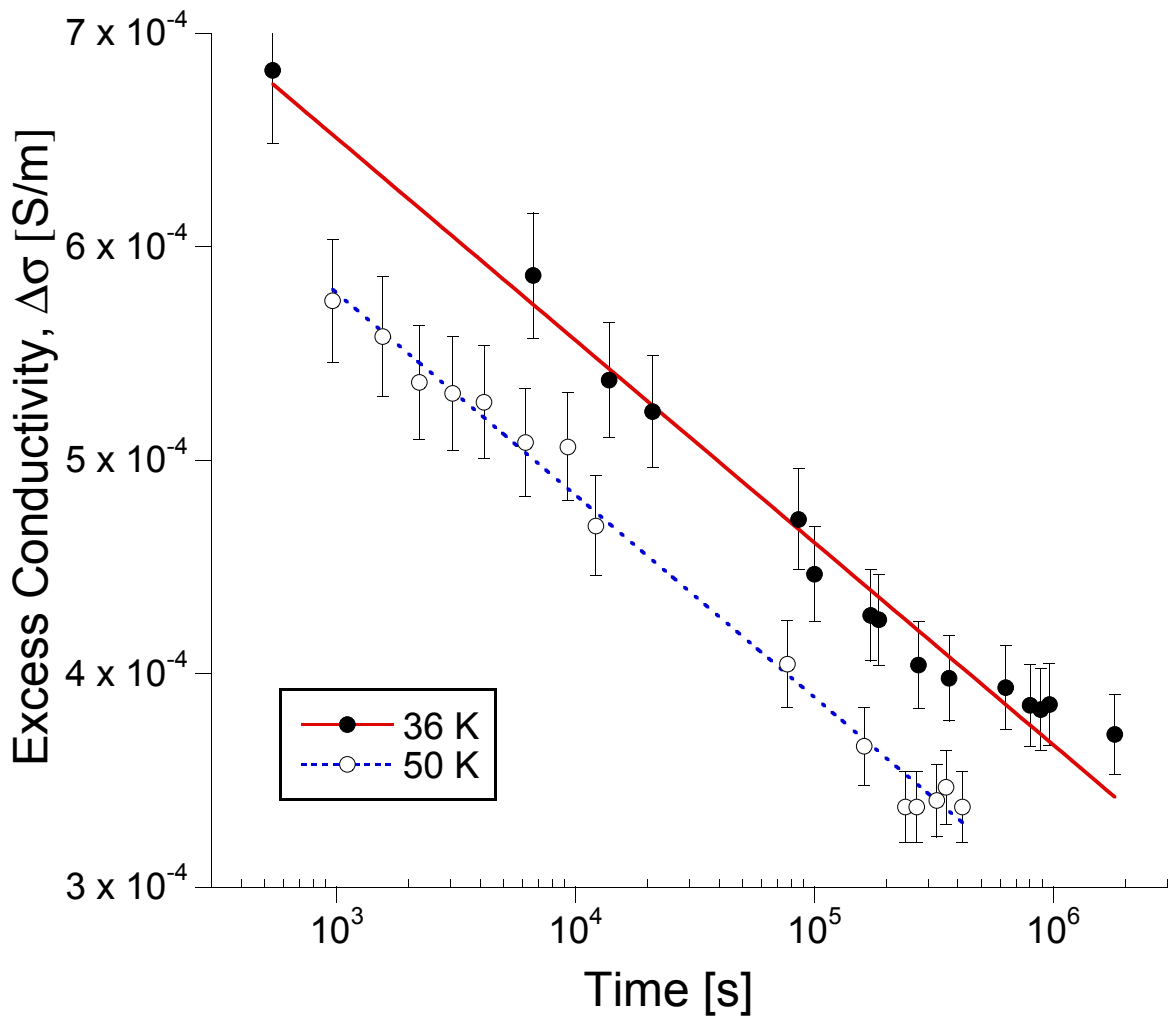
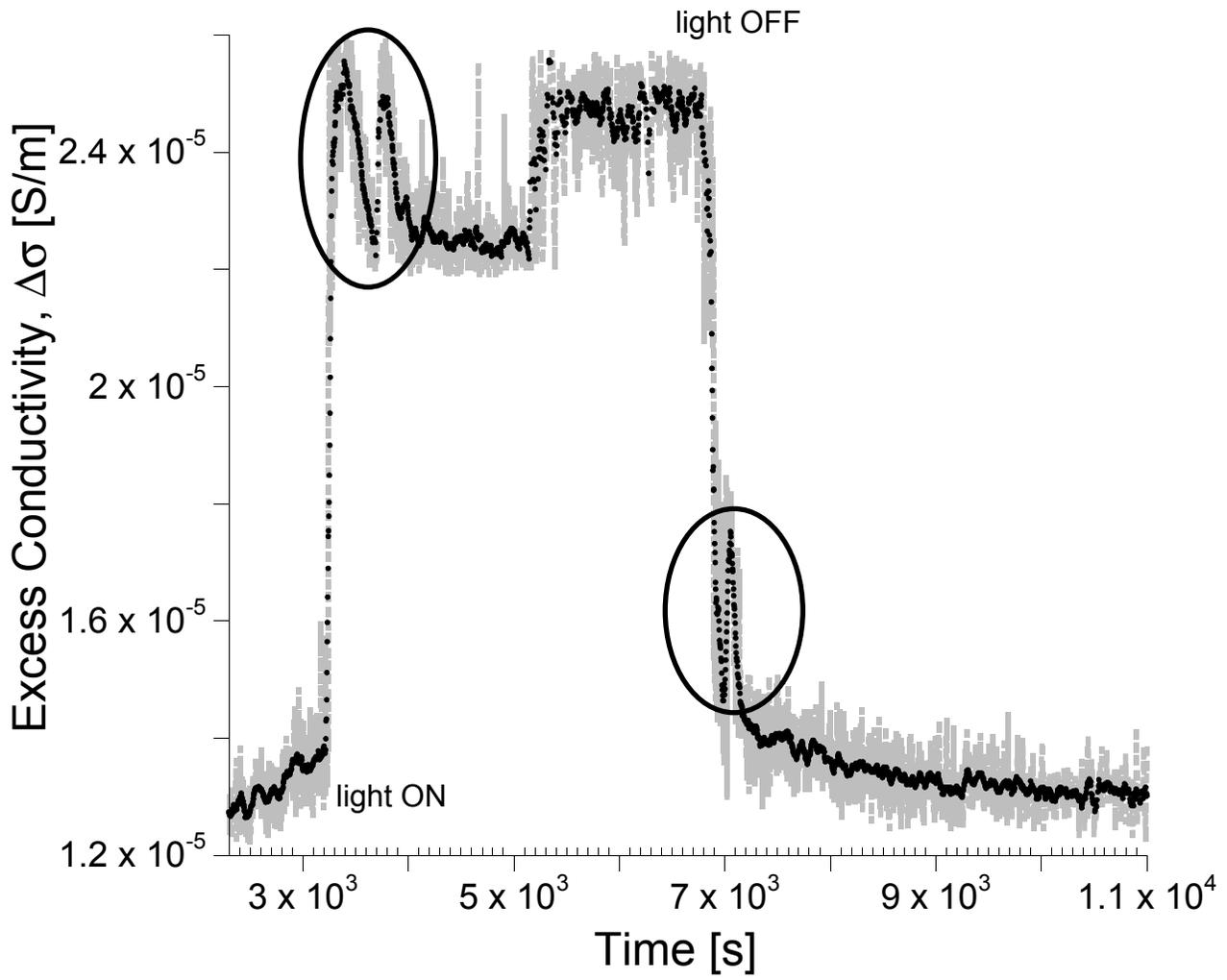
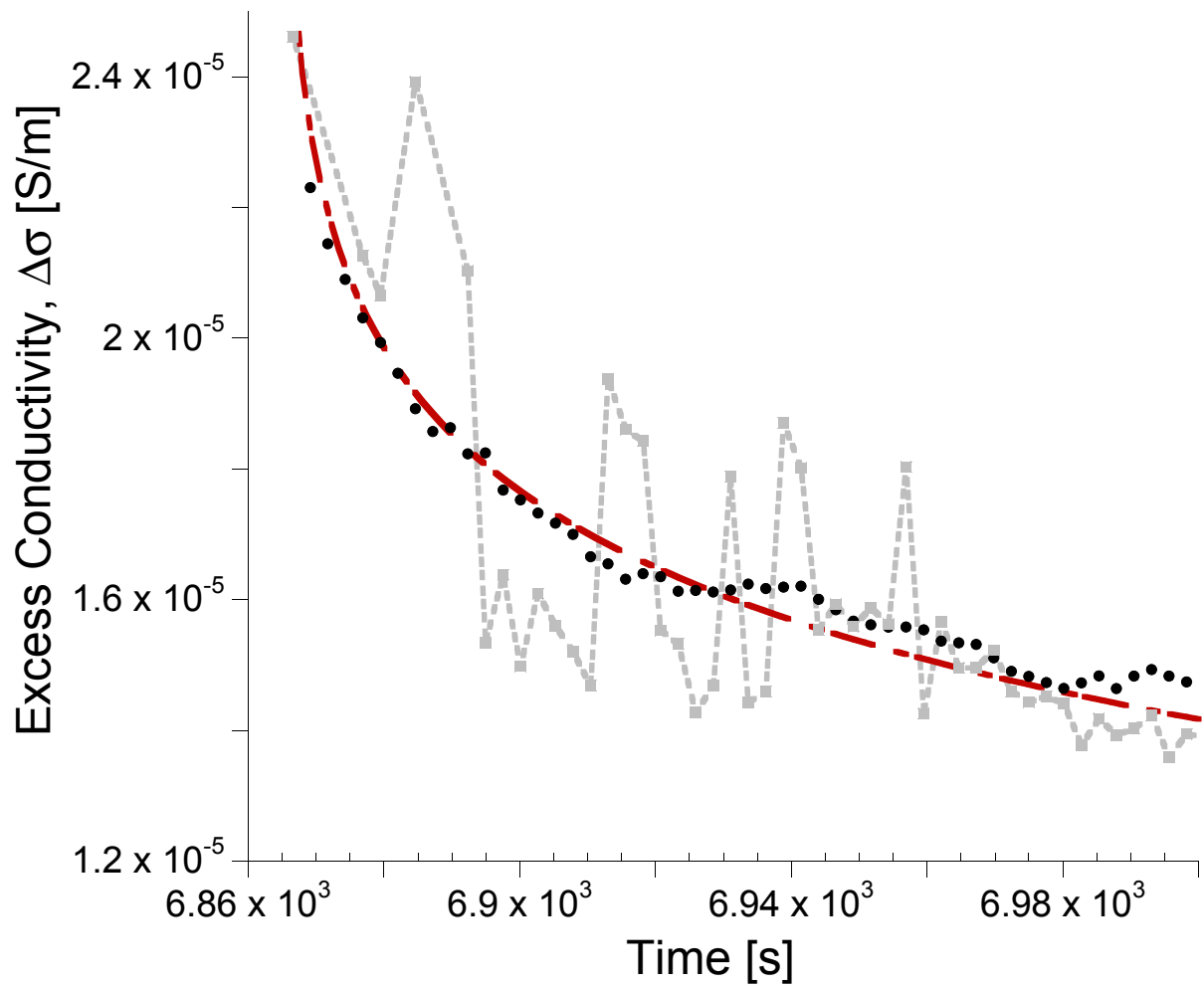
\end{document}